\documentclass[aps,twocolumn,amsfonts,amsmath,amssymb,showkeys,nobibnotes,nofootinbib,floatfix,a4paper]{revtex4-1}

\usepackage{graphicx} 
\usepackage{amsfonts} 
\usepackage{amsmath} 
\usepackage{amssymb} 
\usepackage{hyperref} 
\usepackage{color}
\usepackage{siunitx} 
\usepackage{booktabs} 
\usepackage{hyperref} 
\usepackage{cleveref}
\usepackage{paralist}

\begin{document} 

\title{Basel III capital surcharges for G-SIBs fail to control systemic risk and can cause pro-cyclical side effects} 
\author{Sebastian Poledna$^{1,2}$}
\author{Olaf Bochmann$^{4,5}$}
\author{Stefan Thurner$^{1,2,3}$}
\email{stefan.thurner@meduniwien.ac.at}

\affiliation{
$^1$IIASA, Schlossplatz 1, 2361 Laxenburg; Austria\\
$^2$Section for Science of Complex Systems; Medical University of Vienna; Spitalgasse 23; 1090 Wien; Austria\\
$^3$Santa Fe Institute; 1399 Hyde Park Road; Santa Fe; NM 87501; USA\\
$^4$Institute of New Economic Thinking at the Oxford Martin School, Eagle House, Walton Well Rd., Oxford OX2 3ED; UK\\
$^5$Mathematical Institute, University of Oxford, Woodstock Rd., Oxford OX2 6GG; UK}

\begin{abstract}
In addition to constraining bilateral exposures of financial institutions, there are essentially two options for future financial regulation of systemic risk (SR): First, financial regulation could attempt to reduce the financial fragility of global or domestic systemically important financial institutions (G-SIBs or D-SIBs), as for instance proposed in Basel III. Second, future financial regulation could attempt strengthening the financial system as a whole. This can be achieved by re-shaping the topology of financial networks. We use an agent-based model (ABM) of a financial system and the real economy to study and compare the consequences of these two options. By conducting three ``computer experiments'' with the ABM we find that re-shaping financial networks is more effective and efficient than reducing leverage. Capital surcharges for G-SIBs can reduce SR, but must be larger than those specified in Basel III in order to have a measurable impact. This can cause a loss of efficiency. Basel III capital surcharges for G-SIBs can have pro-cyclical side effects.
\end{abstract}


\maketitle

\section{Introduction} \label{intro}
Six years after the financial crisis of 2007-2008, millions of households worldwide are still struggling to recover from the aftermath of those traumatic events. The majority of losses are indirect, such as people losing homes or jobs, and for the majority, income levels have dropped substantially. For the economy as a whole, and for households and for public budgets, the miseries of the market meltdown of 2007-2008 are not yet over.As a consequence, a consensus for the need for new financial regulation is emerging \cite{Aikman:2013aa}. Future financial regulation should be designed to mitigate risks within the financial system as a whole, and should specifically address the issue of systemic risk (SR). 

SR is the risk that the financial system as a whole, or a large fraction thereof, can no longer perform its function as a credit provider, and as a result collapses. In a narrow sense, it is the notion of contagion or impact from the failure of a financial institution or group of institutions on the financial system and the wider economy \cite{De-Bandt:2000aa,BIS:2010aa}. Generally, it emerges through one of two mechanisms, either through interconnectedness or through the synchronization of behavior of agents (fire sales, margin calls, herding). The latter can be measured by a potential capital shortfall during periods of synchronized behavior where many institutions are simultaneously distressed \cite{Adrian:2011aa,Acharya:2010aa,Brownlees:2012aa,Huang:2012aa}. Measures for a potential capital shortfall are closely related to the leverage of financial institutions \cite{Acharya:2010aa, Brownlees:2012aa}. Interconnectedness is a consequence of the network nature of financial claims and liabilities \cite{Eisenberg:2001aa}. Several studies indicate that financial network measures could potentially serve as early warning indicators for crises \cite{Caballero:2012aa,Billio:2012aa,Minoiu:2013aa}.

In addition to constraining the (potentially harmful) bilateral exposures of financial institutions, there are essentially two options for future financial regulation to address the problem \cite{Haldane:2011aa,Markose:2012aa}: First, financial regulation could attempt to reduce the financial fragility of ``super-spreaders'' or \emph{systemically important financial institutions} (SIFIs), i.e. limiting a potential capital shortfall. This can achieved by reducing the leverage or increasing the capital requirements for SIFIs. ``Super-spreaders'' are  institutions that are either too big, too connected or otherwise too important to fail. However, a reduction of leverage simultaneously reduces efficiency and can lead to pro-cyclical effects \cite{Minsky:1992aa,Fostel:2008aa,Geanakoplos:2010aa,Adrian:2008aa,Brunnermeier:2009aa,Thurner:2012aa,Caccioli:2012aa,Poledna:2014ab,Aymanns:2014aa,Caccioli:2015aa}.
Second, future financial regulation could attempt strengthening the financial system as a whole. It has been noted that different financial network topologies have different probabilities for systemic collapse \cite{Roukny:2013aa}. In this sense the management of SR is reduced to the technical problem of re-shaping the topology of financial networks \cite{Haldane:2011aa}.

The Basel Committee on Banking Supervision (BCBS) recommendation for future financial regulation for SIFIs is an example of the first option. The Basel III framework recognizes SIFIs and, in particular, global and domestic systemically important banks (G-SIBs or D-SIBs). The BCBS recommends increased capital requirements for SIFIs -- the so called ``SIFI surcharges'' \cite{BIS:2010aa,Georg:2011aa}. They propose that SR should be measured in terms of the impact that a bank's failure can have on the global financial system and the wider economy, rather than just the risk that a failure could occur. Therefore they understand SR as a global, system-wide, loss-given-default (LGD) concept, as opposed to a probability of default (PD) concept. Instead of using quantitative models to estimate SR, Basel III proposes an indicator-based approach that includes the size of banks, their interconnectedness, and other quantitative and qualitative aspects of systemic importance. 

There is not much literature on the problem of dynamically re-shaping network topology so that networks adapt over time to function optimally in terms of stability and efficiency. A major problem in NW-based SR management is to provide agents with incentives to re-arrange their local contracts so that global (system-wide) SR is reduced. Recently, it has been noted empirically that individual transactions in the interbank market alter the SR in the total financial system in a measurable way \cite{Poledna:2014aa,Poledna:2015aa}. This allows an estimation of the marginal SR associated with financial transactions, a fact that has been used to propose a tax on systemically relevant transactions \cite{Poledna:2014aa}. It was demonstrated with an agent-based model (ABM) that such a tax -- the {\em systemic risk tax} (SRT) -- is leading to a dynamical re-structuring of financial networks, so that overall SR is substantially reduced \cite{Poledna:2014aa}. 

In this paper we study and compare the consequences of two different options for the regulation of SR with an ABM. As an example for the first option we study Basel III with capital surcharges for G-SIBs and compare it with an example for the second option -- the SRT that leads to a self-organized re-structuring of financial networks. A number of ABMs have been used recently to study interactions between the financial system and the real economy, focusing on destabilizing feedback loops between the two sectors \cite{Delli-Gatti:2009aa,Battiston:2012ab,Tedeschi:2012aa,Porter:2014aa,Thurner:2013aa,Poledna:2014aa,Klimek:2014aa}. We study the different options for the regulation of SR within the framework of the CRISIS macro-financial model\footnote{\url{http://www.crisis-economics.eu}}. In this ABM, we implement both the Basel III indicator-based measurement approach, and the increased capital requirements for G-SIBs. We compare both to an implementation of the SRT developed in \cite{Poledna:2014aa}. We conduct three ``computer experiments'' with the different regulation schemes. First, we investigate which of the two options to regulate SR is superior. Second, we study the effect of increased capital requirements, the ``surcharges'', on G-SIBs and the real economy. Third, we clarify to what extend the Basel III indicator-based measurement approach really quantifies SR, as indented by the BCBS.

\section{Basel III indicator-based measurement approach and capital surcharges for G-SIBs}
\subsection{Basel III indicator-based measurement approach}
The Basel III indicator-based measurement approach consists of five broad categories: size, interconnectedness, lack of readily available substitutes or financial institution infrastructure, global (cross-jurisdictional) activity and ``complexity''. As shown in \cref{indicator}, the measure gives equal weight to each of the five categories. Each category may again contain individual indicators, which are equally weighted within the category.
\begin{table*}
	\begin{center}
		\begin{tabular}
			{p{5cm} p{7cm} p{3cm}} Category (and weighting) & Individual indicator & Indicator weighting\\
			\hline Cross-jurisdictional activity (20\%) & Cross-jurisdictional claims & 10\%\\
			& Cross-jurisdictional liabilities & 10\%\\
			Size (20\%) & Total exposures as defined for use in the Basel III leverage ratio & 20\%\\
			Interconnectedness (20\%) & Intra-financial system assets & 6.67\%\\
			& Intra-financial system liabilities & 6.67\%\\
			& Securities outstanding & 6.67\%\\
			Substitutability/financial institution infrastructure (20\%) & Assets under custody & 6.67\%\\
			& Payments activity & 6.67\%\\
			& Underwritten transactions in debt and equity markets & 6.67\%\\
			Complexity (20\%) & Notional amount of over-the-counter (OTC) derivatives & 6.67\%\\
			& Level 3 assets & 6.67\%\\
			& Trading and available-for-sale securities & 6.67\% 
		\end{tabular}
	\end{center}
	\caption{Basel III indicators and indicator weights for the ``indicator-based measurement approach''.} \label{indicator} 
\end{table*}
Here below we describe each of the categories in more detail. 
\begin{description}
	\item[Cross-jurisdiction activity] This indicator captures the global ``footprint'' of Banks. The motivation is to reflect the coordination difficulty associated with the resolution of international spillover effects. It measures the bank's cross-jurisdictional activity relative to other banks activities. Here it differentiates between 
	\begin{inparaenum}
		[(i)] 
		\item cross-jurisdictional claims, and 
		\item cross-jurisdictional liabilities. 
	\end{inparaenum}
	
	\item[Size] This indicator reflects the idea that size matters. Larger banks are more difficult to replace and the failure of a large bank is more likely to damage confidence in the system. The indicator is a measure of the normalized total exposure used in Basel III leverage ratio.
	
	\item[Interconnectedness] A bank's systemic impact is likely to be related to its interconnectedness to other institutions via a network of contractual obligations. The indicator differentiates between 
	\begin{inparaenum}
		[(i)] 
		\item in-degree on the network of contracts 
		\item out-degree on the network of contracts, and 
		\item outstanding securities. 
	\end{inparaenum}
	
	\item[Substitutability/financial institution infrastructure] The indicator for this category captures the increased difficulty in replacing banks that provide unique services or infrastructure. The indicator differentiates between 
	\begin{inparaenum}
		[(i)] 
		\item assets under custody 
		\item payment activity, and 
		\item underwritten transactions in debt and equity markets. 
	\end{inparaenum}
	
	\item[Complexity] Banks are complex in terms of business structure and operational ``complexity''. The costs for resolving a complex bank is considered to be greater. This is reflected in the indicator as 
	\begin{inparaenum}
		[(i)] 
		\item notional amount of over-the-counter derivatives 
		\item level 3 assets, and 
		\item trading and available-for-sale securities. 
	\end{inparaenum}
\end{description}

The score $S_{j}$ of the Basel III indicator-based measurement approach for each bank $j$ and each indicator $D^{i}$, e.g. cross-jurisdictional claims, is calculated as the fraction of the individual banks with respect to all $B$ Banks and then weighted by the indicator weight ($\beta_i$). The score is given in basis points (factor $10000$) 
\begin{equation}
	S_{j} = \sum_{i \in I} \beta_i \frac{D^{i}_{j}}{\sum_j^B D^{i}_{j}}10000 \quad, \label{score}
\end{equation}
where $I$ is the set of indicators $D^{i}$ and $\beta_i$ the weights from \cref{indicator}.

\subsection{Basel III capital surcharges for G-SIBs}
\begin{table*}
	\begin{center}
		\begin{tabular}
			{p{3cm} p{3cm} p{3cm} p{6cm}} Bucket & Score range & Bucket thresholds & Higher loss absorbency requirement (common equity as a percentage of risk-weighted assets)\\
			\hline 5 & D-E & 530-629 & 3.50\%\\
			4 & C-D & 430-529 & 2.50\%\\
			3 & B-C & 330-429 & 2.00\%\\
			2 & A-B & 230-329 & 1.50\%\\
			1 & Cutoff point-A & 130-229 & 1.00\% 
		\end{tabular}
	\end{center}
	\caption{Categorization of systemic importance in the ``bucketing approach'', as proposed by the BCBS.} \label{buckets} 
\end{table*}

In the Basel III ``bucketing approach'', based on the scores from \cref{score}, banks are divided into four equally sized classes (buckets) of systemic importance, seen here in \cref{buckets}. The cutoff score and bucket thresholds have been calibrated by the BCBS in such a way that the magnitude of the higher loss absorbency requirements for the highest populated bucket is 2.5\% of risk-weighted assets, with an initially empty bucket of 3.5\% of risk-weighted assets. The loss absorbency requirements for the lowest bucket is 1\% of risk-weighted assets. The loss absorbency requirement is to be met with common equity \cite{BIS:2010aa}. Bucket five will initially be empty. As soon as the bucket becomes populated, a new bucket will be added in such a way that it is equal in size (scores) to each of the other populated buckets and the minimum higher loss absorbency requirement is increased by 1\% of risk-weighted assets.

\section{The model to test the efficiency of SR regulation} \label{model}
We use an ABM, linking the financial and the real economy. The model consists of banks, firms and households. One pillar of the model is a well-studied macroeconomic model \cite{Gaffeo:2008aa,Delli-Gatti:2008aa,Delli-Gatti:2011aa}, the second pillar is an implementation of an interbank market. In particular, we extend the model used in \cite{Poledna:2014aa} with an implementation of the Basel III indicator-based measurement approach and capital surcharges for SIBs. For a comprehensive description of the model, see \cite{Delli-Gatti:2011aa,Gualdi:2013aa,Poledna:2014aa}. 

The agents in the model interact on four different markets. 
\begin{inparaenum}
	[(i)]
	\item Firms and banks interact on the credit market, generating flows of loan (re)payments. 
	\item Banks interact with other banks on the interbank market, generating flows of interbank loan (re)payments. 
	\item Households and firms interact on the labour market, generating flows wage payments.
	\item Households and firms interact on the consumption-goods market, generating flows of goods. 
\end{inparaenum}
Banks hold all of firms and households? cash as deposits. Households are randomly assigned as owners of firms and banks (share-holders). 

Agents repeat the following sequence of decisions at each time step:  
\begin{inparaenum}
	\item firms define labour and capital demand,
	\item banks rise liquidity for loans,
	\item firms allocate capital for production (labour),
	\item households receive wages, decide on consumption and savings,
	\item firms and banks pay dividends, firms with negative cash go bankrupt,
	\item banks and firms repay loans,
	\item illiquid banks try to rise liquidity, and if unsuccessful, go bankrupt. 
\end{inparaenum}
Households which own firms or banks use dividends as income, all other households use wages. Banks and firms pay $20\%$ of their profits as dividends. 
The agents are described in more detail below.

\subsubsection{Households} 
There are $H$ households in the model. Households can either be workers or investors that own firms or banks. Each household $j$ has a personal account $A_{j,b}(t)$ at one of the $B$ banks, where index $j$ represents the household and index $b$ the bank. Workers apply for jobs at the $F$ different firms. If hired, they receive a fixed wage $w$ per time step, and supply a fixed labour productivity $\alpha$. A household spends a fixed percentage $c$ of its current account on the consumption goods market. They compare prices of goods from $z$ randomly chosen firms and buy the cheapest.

\subsubsection{Firms} 
There are $F$ firms in the model. They produce perfectly substitutable goods. At each time step firm $i$ computes its own expected demand $d_i(t)$ and price $p_i(t)$. The estimation is based on a rule that takes into account both excess demand/supply and the deviation of the price $p_i(t-1)$ from the average price in the previous time step \cite{Delli-Gatti:2011aa}. Each firm computes the required labour to meet the expected demand. If the wages for the respective workforce exceed the firm's current liquidity, the firm applies for a loan. Firms approach $n$ randomly chosen banks and choose the loan with the most favorable rate. If this rate exceeds a threshold rate $r^{\rm max}$, the firm only asks for $\phi$ percent of the originally desired loan volume. Based on the outcome of this loan request, firms re-evaluate the necessary workforce, and hire or fire accordingly. Firms sell the produced goods on the consumption goods market. Firms go bankrupt if they run into negative liquidity. Each of the bankrupted firms' debtors (banks) incurs a capital loss in proportion to their investment in loans to the company. Firm owners of bankrupted firms are personally liable. Their account is divided by the debtors \emph{pro rata}. They immediately (next time step) start a new company, which initially has zero equity. Their initial estimates for demand  $d_i(t)$ and price $p_i(t)$ is set to the respective averages in the goods market. 

\subsubsection{Banks} 
The model involves $B$ banks. Banks offer firm loans at rates that take into account the individual specificity of banks (modeled by a uniformly distributed random variable), and the firms' creditworthiness. Firms pay a credit risk premium according to their creditworthiness that is modeled by a monotonically increasing function of their financial fragility \cite{Delli-Gatti:2011aa}. Banks try to grant these requests for firm loans, providing they have enough liquid resources. If they do not have enough cash, they approach other banks in the interbank market to obtain the necessary amount. If a bank does not have enough cash and can not raise the total amount requested on the interbank market, it does not pay out the loan. Interbank and firm loans have the same duration. Additional refinancing costs of banks remain with the firms. Each time step firms and banks repay $\tau$ percent of their outstanding debt (principal plus interest). If banks have excess liquidity they offer it on the interbank market for a nominal interest rate. The interbank relation network is modeled as a fully connected network and banks choose the interbank offer with the most favorable rate. Interbank rates $r_{ij}(t)$ offered by bank $i$ to bank $j$ take into account the specificity of bank $i$, and the creditworthiness of bank $j$. If a firm goes bankrupt the respective creditor bank writes off the respective outstanding loans as defaulted credits. If the bank does not have enough equity capital to cover these losses it defaults. Following a bank default an iterative default-event unfolds for all interbank creditors. This may trigger a cascade of bank defaults. For simplicity's sake, we assume a recovery rate set to zero for interbank loans. This assumption is reasonable in practice for short term liquidity \cite{Cont:2013aa}. A cascade of bankruptcies happens within one time step. After the last bankruptcy is resolved the simulation is stopped.

\subsection{Implementation of the Basel III indicator-based measurement approach}
In the ABM, we implement the size indicator by calculating the total exposures of banks. Total exposure includes all assets of banks excluding cash, i.e. loans to firms and loans to other banks. Interconnectedness is measured in the model by interbank assets (loans) and interbank liabilities (deposits) of banks. As a measure for substitutability we use the payment activity of banks. The payment activity is measured by the sum of all outgoing payments by banks. In the model we do not have cross-jurisdictional activity and banks are not engaged in selling complex financial products including derivatives and level 3 assets. We therefore set the weights for global (cross-jurisdictional) activity and complexity to zero. 

Banks in the model have to observe loss absorbency (capital) requirements according to Basel III. They are required to hold 4.5\% of common equity (up from 2\% in Basel II) of risk-weighted assets (RWAs).  RWAs are calculated according to the standardized approach, i.e. with fixed weights for all asset classes. As fixed weights we use 100\% for interbank loans and commercial loans. We define equity capital of banks in the model as common equity. 

In the model banks are allocated to the five buckets shown in \cref{buckets} based on the scores obtained by \cref{score}. Banks have to meet additional loss absorbency requirements as shown in \cref{buckets}. The score is calculated at every time step $t$ and capital requirements must be observed before providing new loans. 

\subsection{Implementation of the SRT in the model}
The SRT is implemented in the model as described in \cite{Poledna:2014aa}, for details see \cref{srt} and \cite{Poledna:2014aa}. The SRT is calculated according to \cref{srteq_simple} and is imposed on all interbank transactions. All other transactions are exempted from SRT. Before entering a desired loan contraction, loan requesting bank $i$ obtains quotes for the SRT rates from the Central Bank, for various potential lenders $j$. Bank $i$ chooses the interbank offer from bank $j$ with the smallest total rate. The SRT is collected into a bailout fund. 

\section{Results} \label{results}
Below we present the results of three experiments. The first experiment focuses on the performance of different options for regulation of SR, the second experiment on the consequences of different capital surcharges for G-SIBs, and the third experiment on the effects of different weight distributions for the Basel III indicator-based measurement approach. 

\subsection{Performance of different options for regulation of SR}
\begin{figure*}
	\begin{center}
		\includegraphics[width=.49\textwidth]{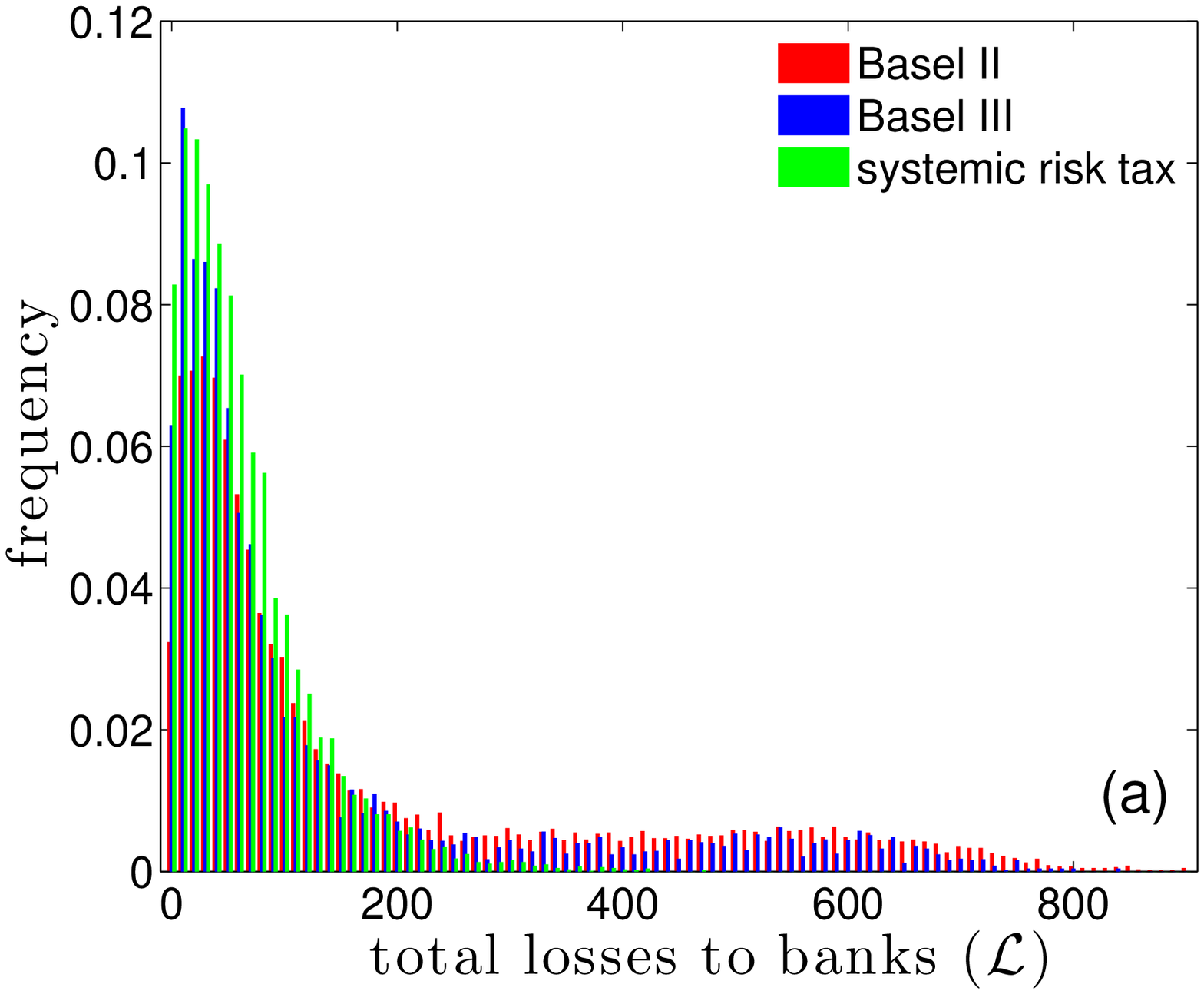}
		\includegraphics[width=.49\textwidth]{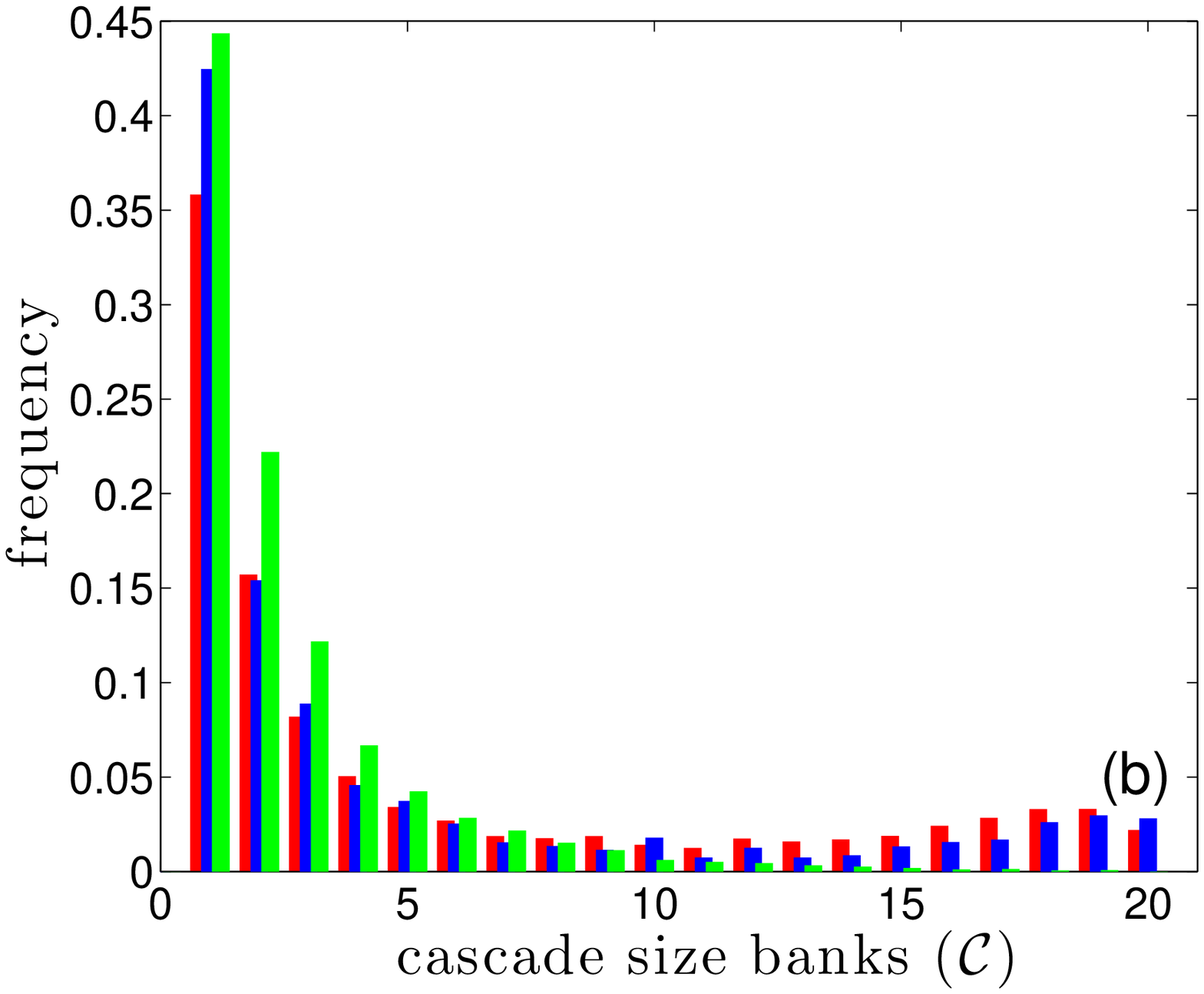}
		\includegraphics[width=.49\textwidth]{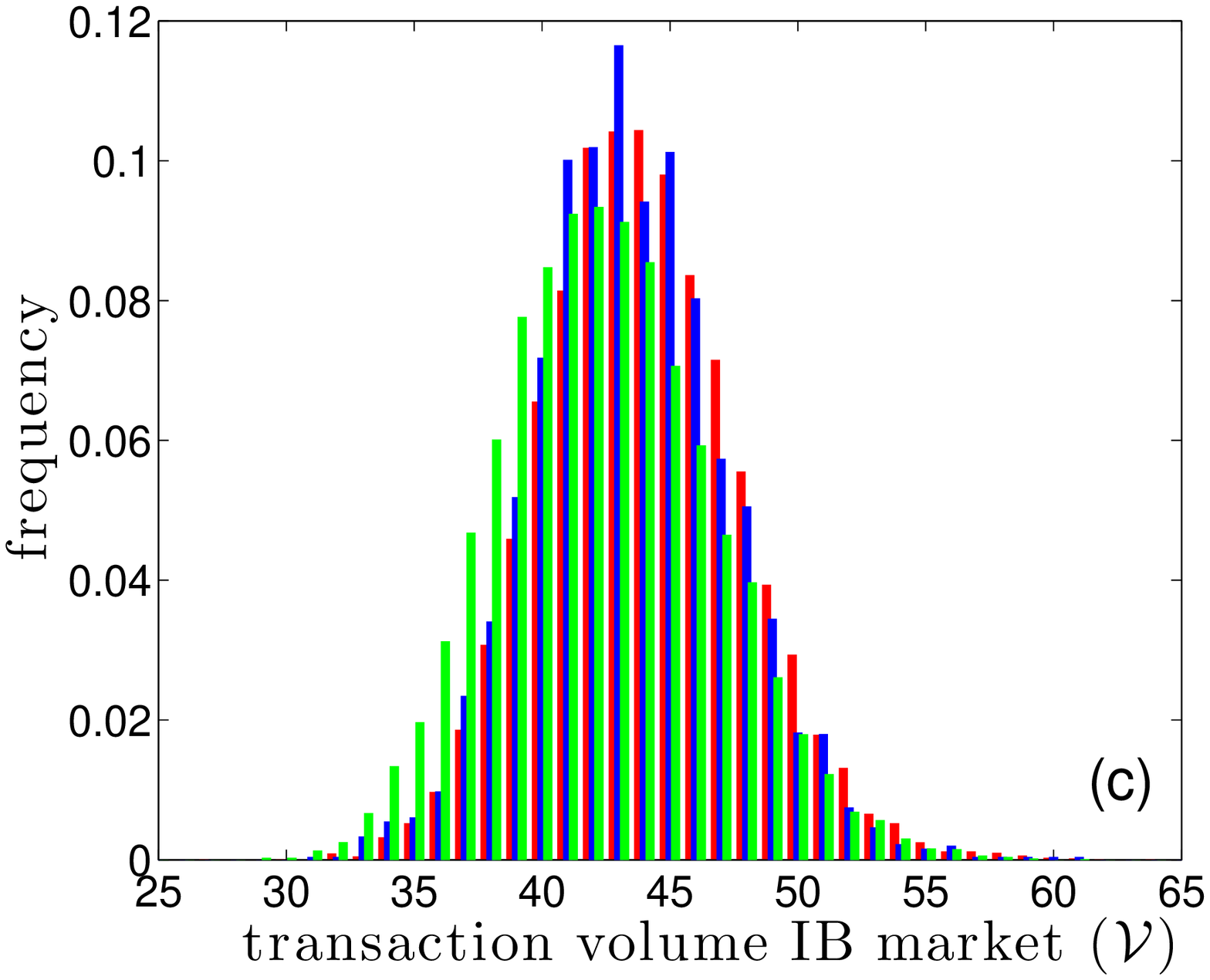}
		\includegraphics[width=.49\textwidth]{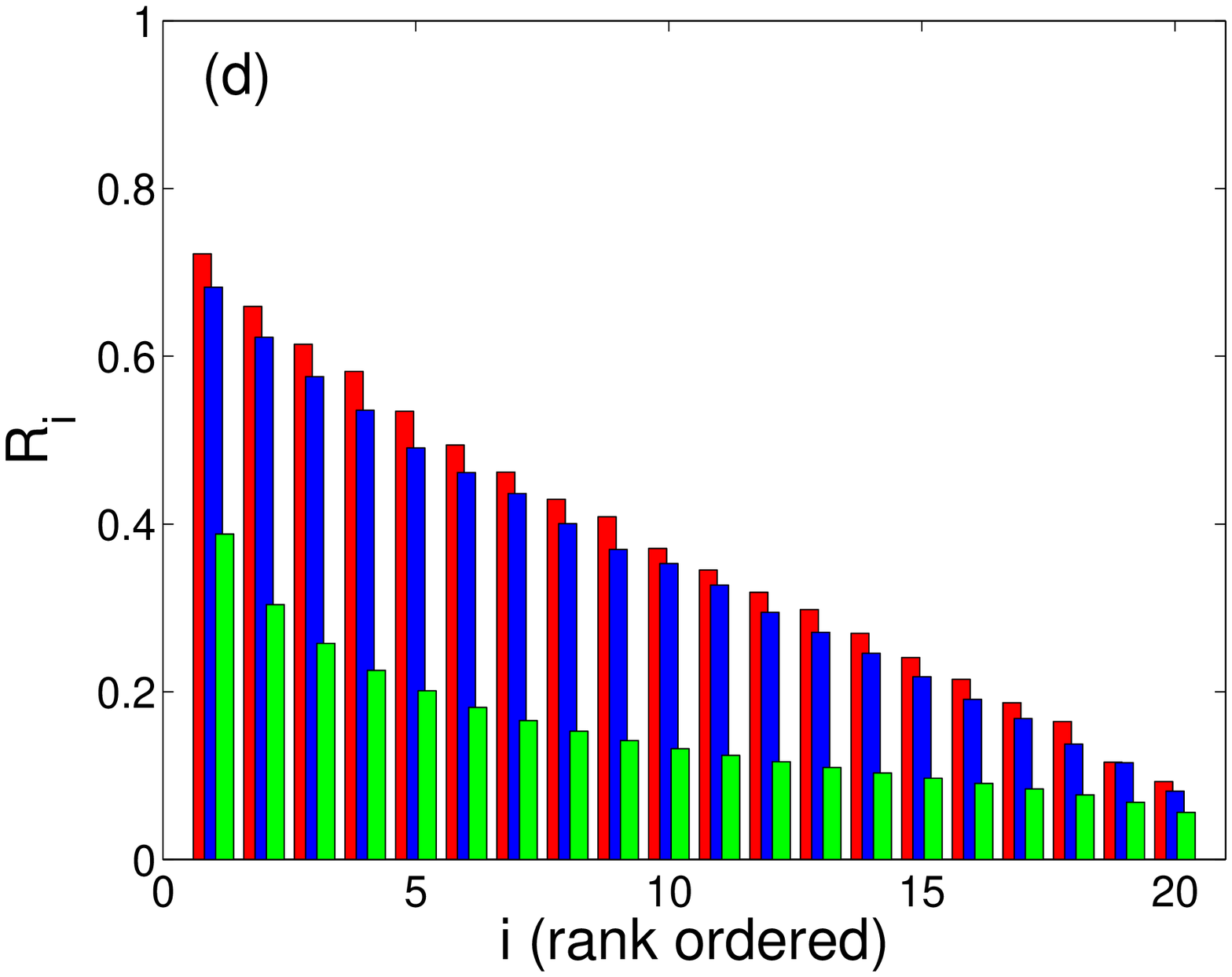}
	\end{center}
	\caption{Comparison of different options for regulation of systemic risk: Basel II without mitigation of systemic risk (red), Basel III with capital surcharges for G-SIBs (blue), and the systemic risk tax (green). (a) Distribution of total losses to banks ${\cal L}$, (b) distribution of cascade sizes ${\cal C}$ of defaulting banks, (c) distribution of total transaction volume in the interbank market ${\cal V}$, and (d) the risk profile for systemic risk given by the distribution of DebtRanks $R_{i}$ (banks are ordered by $R_{i}$, the most systemically important being to the very left). Basel III with capital surcharges for G-SIBs only slightly reduces cascading failure (barely visible), while the systemic risk tax almost gets completely rid of big losses in the system. Efficiency (credit volume) is not affected by the different regulation policies.}
	\label{fig:regulation_type}
\end{figure*}
In order to understand the effects of Basel III capital surcharges for G-SIBs on the financial system and the real economy, and to compare it with the SRT, we conduct an experiment where we consider three cases:
\begin{inparaenum}
	[(i)]
	\item financial system regulated with Basel II,
	\item financial system regulated with Basel III, with capital surcharges for G-SIBs,
	\item financial system with the SRT.
\end{inparaenum}
With Basel II we require banks to hold 2\% of common equity of their RWAs. RWAs are calculated according to the standardized approach. 

The effects of the different regulation policies on total losses to banks ${\cal L}$ (see \cref{risk_measures}) are shown in \cref{fig:regulation_type}(a). Clearly, with Basel II (red) fat tails in the loss distributions of the banking sector are visible. Basel III with capital surcharges for G-SIBs slightly reduces the losses (almost not visible). The SRT gets almost completely rid of big losses in the system (green). This reduction in major losses in the financial system is due to the fact that with the SRT the possibility of cascading failure is drastically reduced. This is shown in \cref{fig:regulation_type}(b) where the distributions of cascade sizes $\mathcal{S}$ (see \cref{risk_measures}) for the three modes are compared. With the Basel regulation policies we observe cascade sizes of up to $20$ banks, while with the SRT the possibility of cascading failure is drastically reduced. Clearly, the total transaction volume $\mathcal{V}$ (see \cref{risk_measures}) in the interbank market is not affected by the different regulation policies. This is seen in \cref{fig:regulation_type}(c), where we show the distributions of transaction volumes. 

In \cref{fig:regulation_type}(d) we show the {\em SR-profile} of the financial system given by the rank-ordered DebtRanks $R_i$ of all banks \citep{Poledna:2015aa}. The SR-profile shows the distribution of systemic impact across banks in the financial system. The bank with the highest systemic impact is to the very left. Obviously, the SRT drastically reduces the systemic impact of individual banks and leads to a more homogeneous spreading of SR across banks. Basel III with capital surcharges for G-SIBs sees a slight reduction in the systemic impact of individual banks.  

\subsection{Consequences of different capital surcharges for G-SIBs}
\begin{figure*}
	\begin{center}
		\includegraphics[width=.49\textwidth]{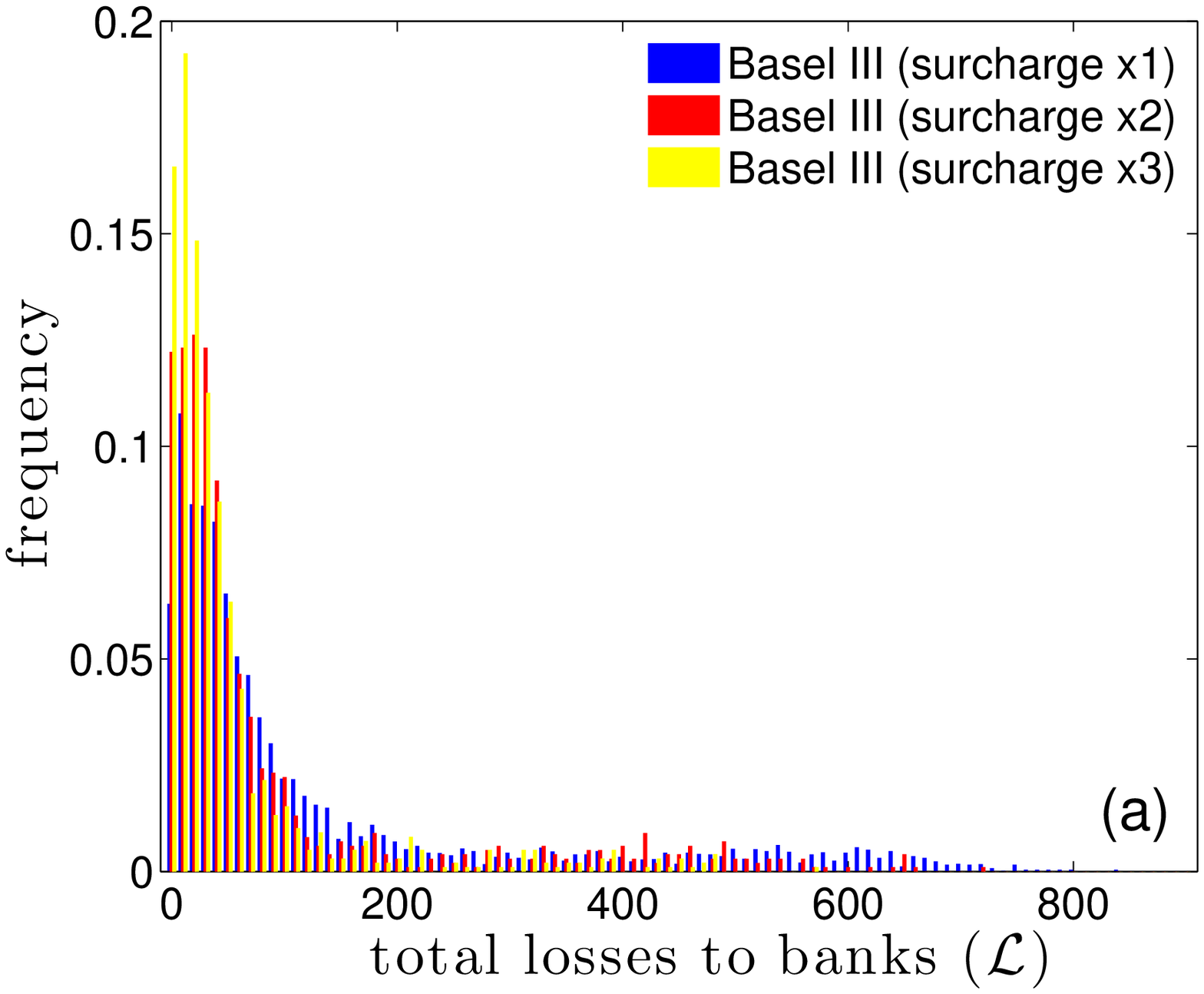}
		\includegraphics[width=.49\textwidth]{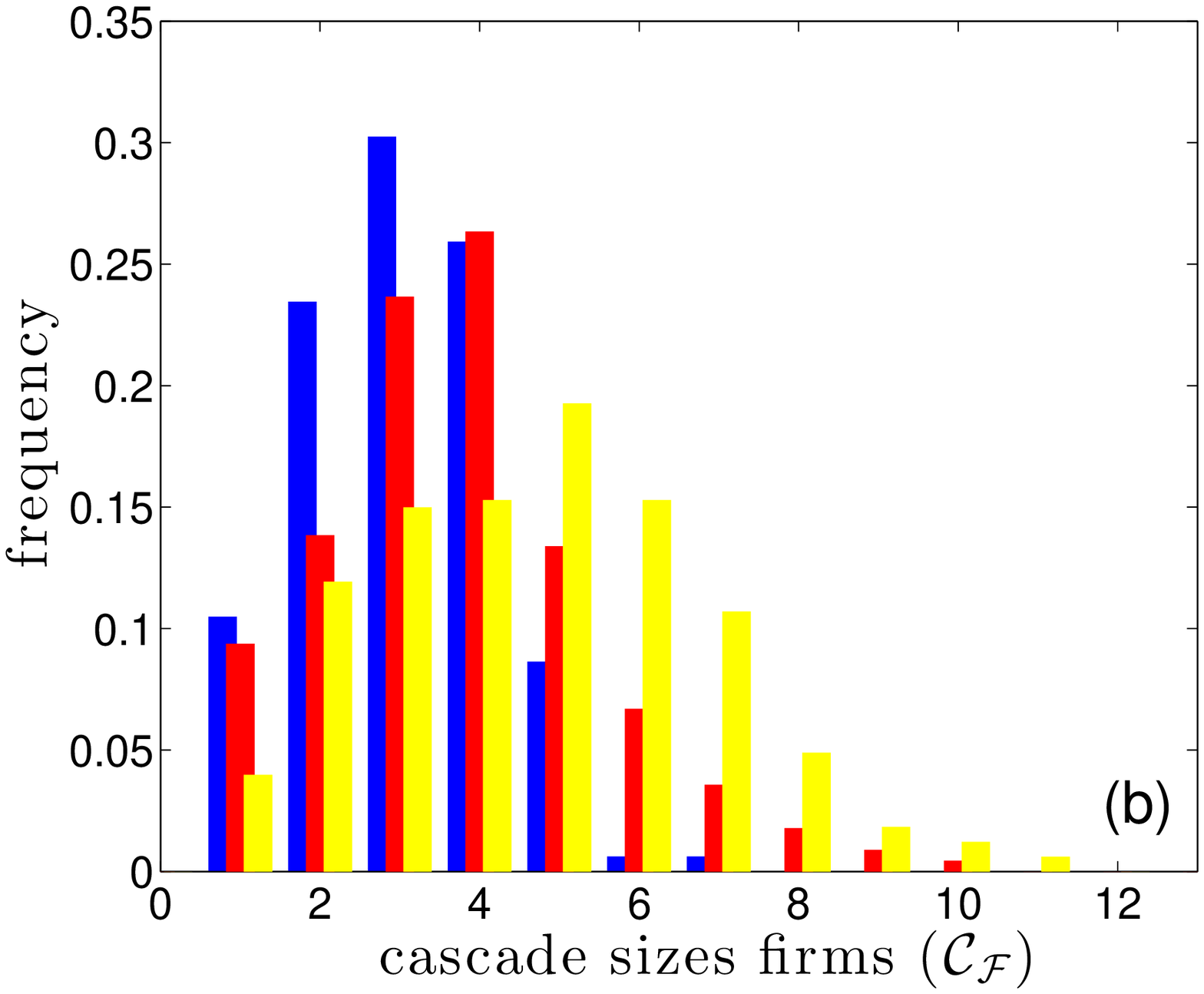}
		\includegraphics[width=.49\textwidth]{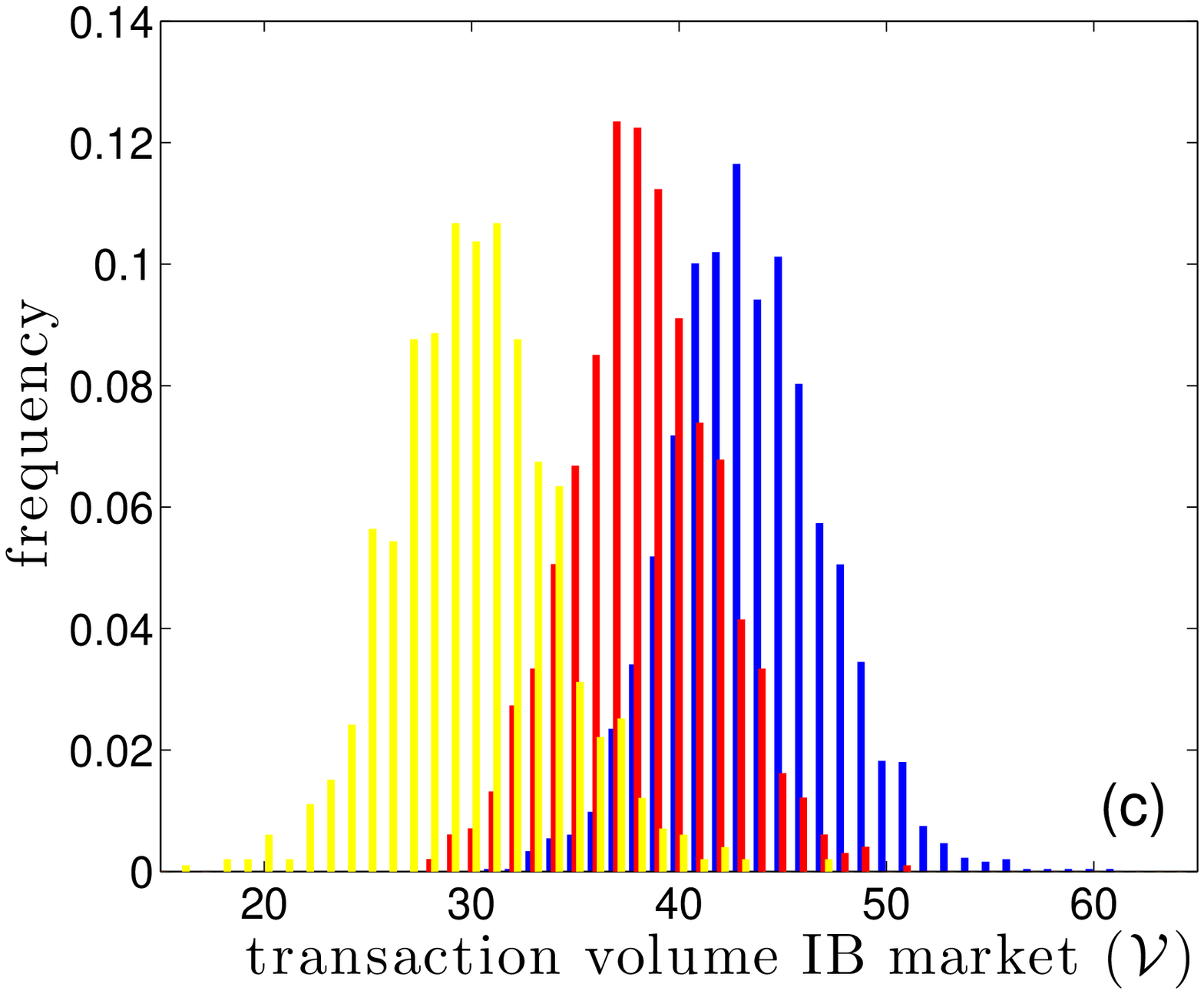}
		\includegraphics[width=.49\textwidth]{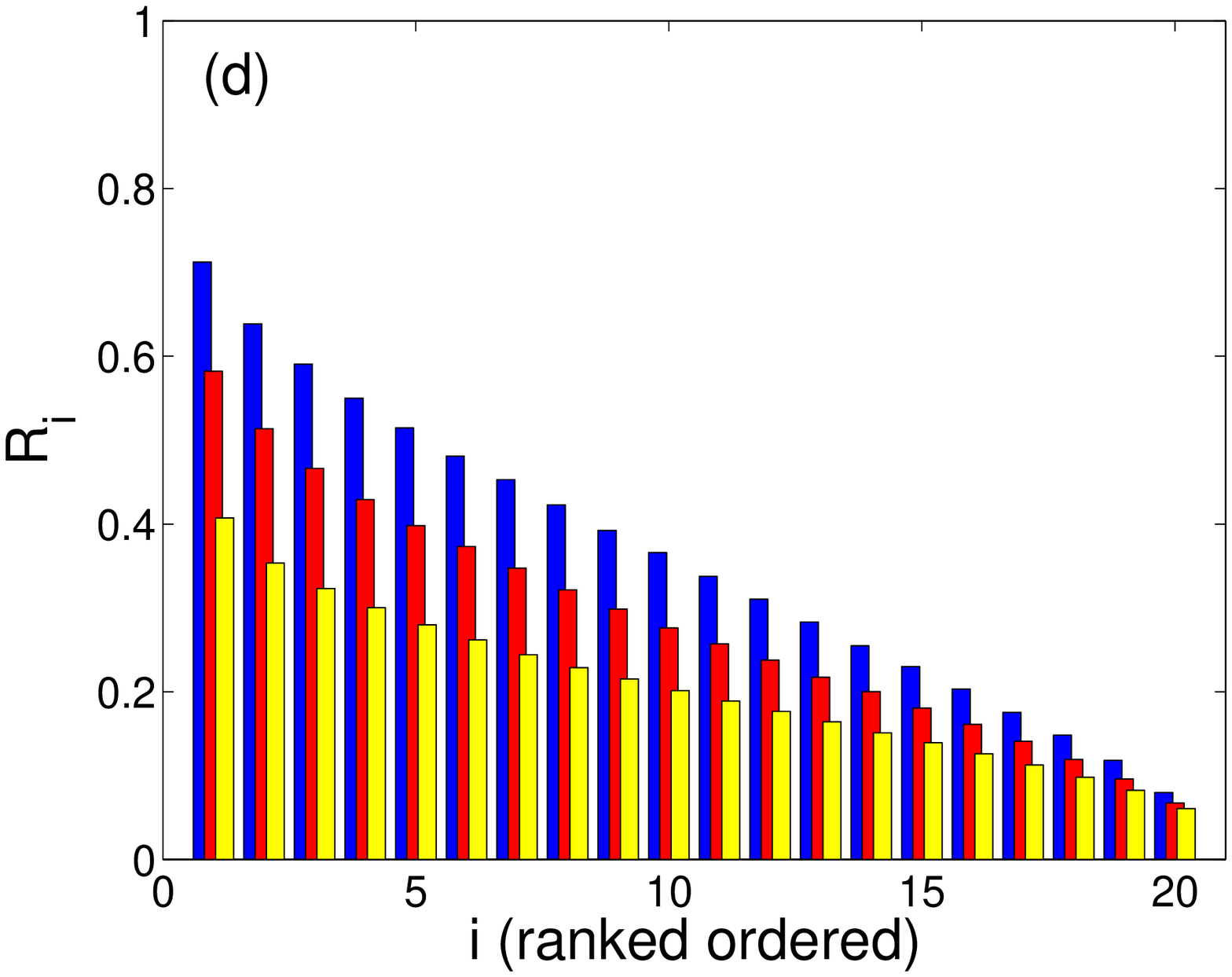}
	\end{center}
	\caption{Comparison of different levels of capital surcharges for G-SIBs: surcharge x1 (blue), surcharge x2 (red), and surcharge x3 (yellow). (a) Distribution of total losses to banks ${\cal L}$, (b) distribution of cascade sizes of synchronized firm defaults, (c) distribution of total transaction volume in the interbank market ${\cal V}$, and (d) the risk profile for systemic risk given by the distribution of DebtRanks $R_{i}$ (banks are ordered by $R_{i}$, the most systemically important being to the very left). With larger capital surcharges for G-SIBs, Basel III capital surcharges have a more visible effect on the financial system. Average losses ${\cal L}$ are reduced at the cost of a loss of efficiency by roughly the same factor.}
	\label{fig:surcharges}
\end{figure*}
Next, we study the effect of different levels of capital surcharges for G-SIBs. Here we consider three different settings:
\begin{inparaenum}
	[(i)]
	\item capital surcharges for G-SIBs as specified in Basel III (\cref{buckets}),
	\item double capital surcharges for G-SIBs,
	\item threefold capital surcharges for G-SIBs.
\end{inparaenum}
With larger capital surcharges for G-SIBs, we observe a stronger effect of the Basel III regulation policy on the financial system. Clearly, the shape of the distribution of losses ${\cal L}$ is similar (\cref{fig:surcharges}(a)). The tail of the distributions is only reduced due to a decrease in efficiency (transaction volume), as is seen in \cref{fig:surcharges}(c). Evidently, average losses ${\cal L}$ are reduced at the cost of a loss of efficiency by roughly the same factor. This means that Basel III must reduce efficiency in order to show any effect in terms of SR, see \cref{discussion}.

In \cref{fig:surcharges}(d) we show the SR-profile for different capital surcharges for G-SIBs. Clearly, the systemic impact of individual banks is also reduced at the cost of a loss of efficiency by roughly the same factor.

In \cref{fig:surcharges}(b) we show the cascade size of synchronized firm defaults. Here we see an increase in cascading failure of firms with increasing capital surcharges for G-SIBs. This means larger capital surcharges for G-SIBs can have pro-cyclical side effects, see \cref{discussion}.

\subsection{Does the Basel III indicator-based measurement approach measure SR?}
\begin{figure*}
	\begin{center}
		\includegraphics[width=.49\textwidth]{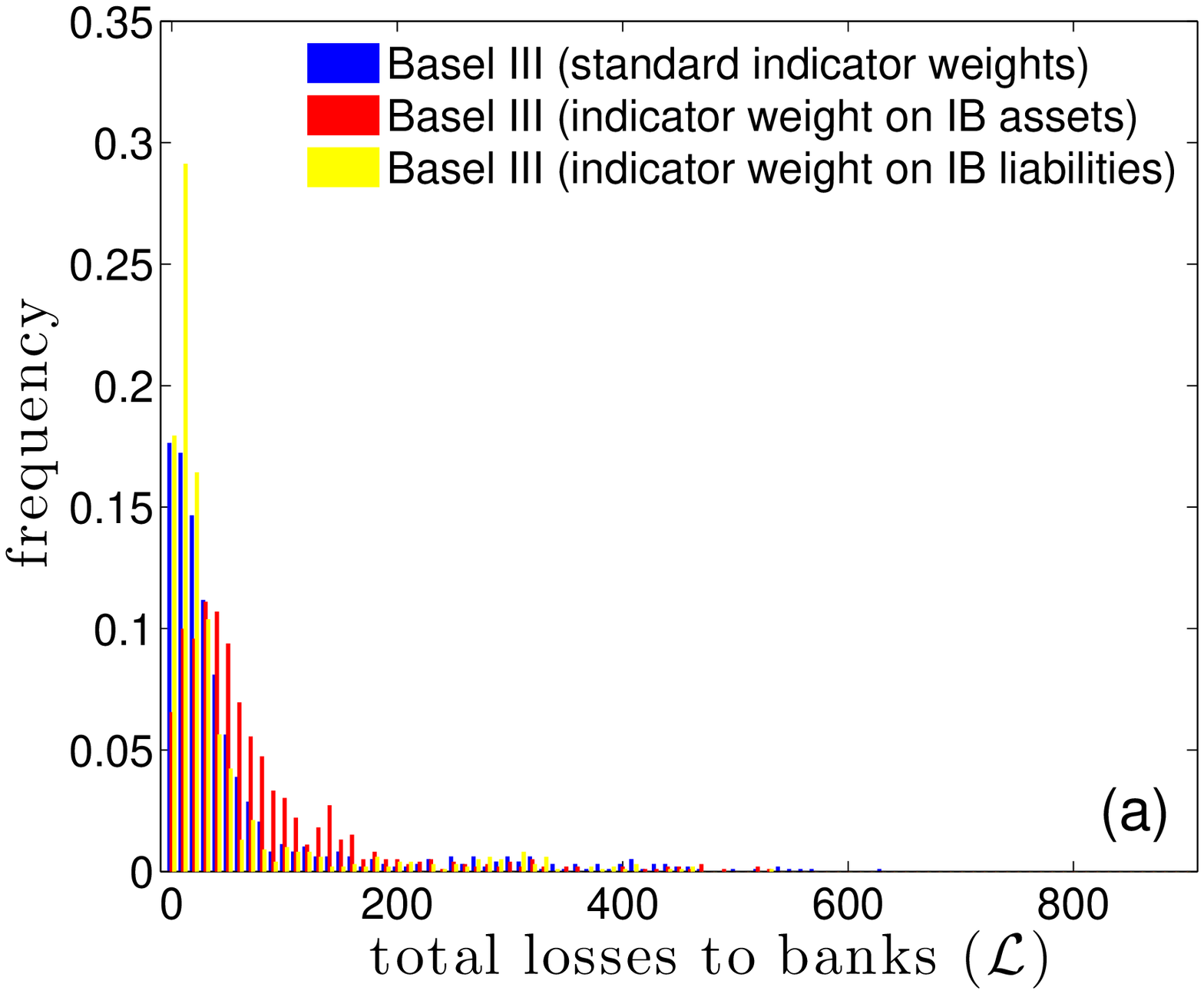}
		\includegraphics[width=.49\textwidth]{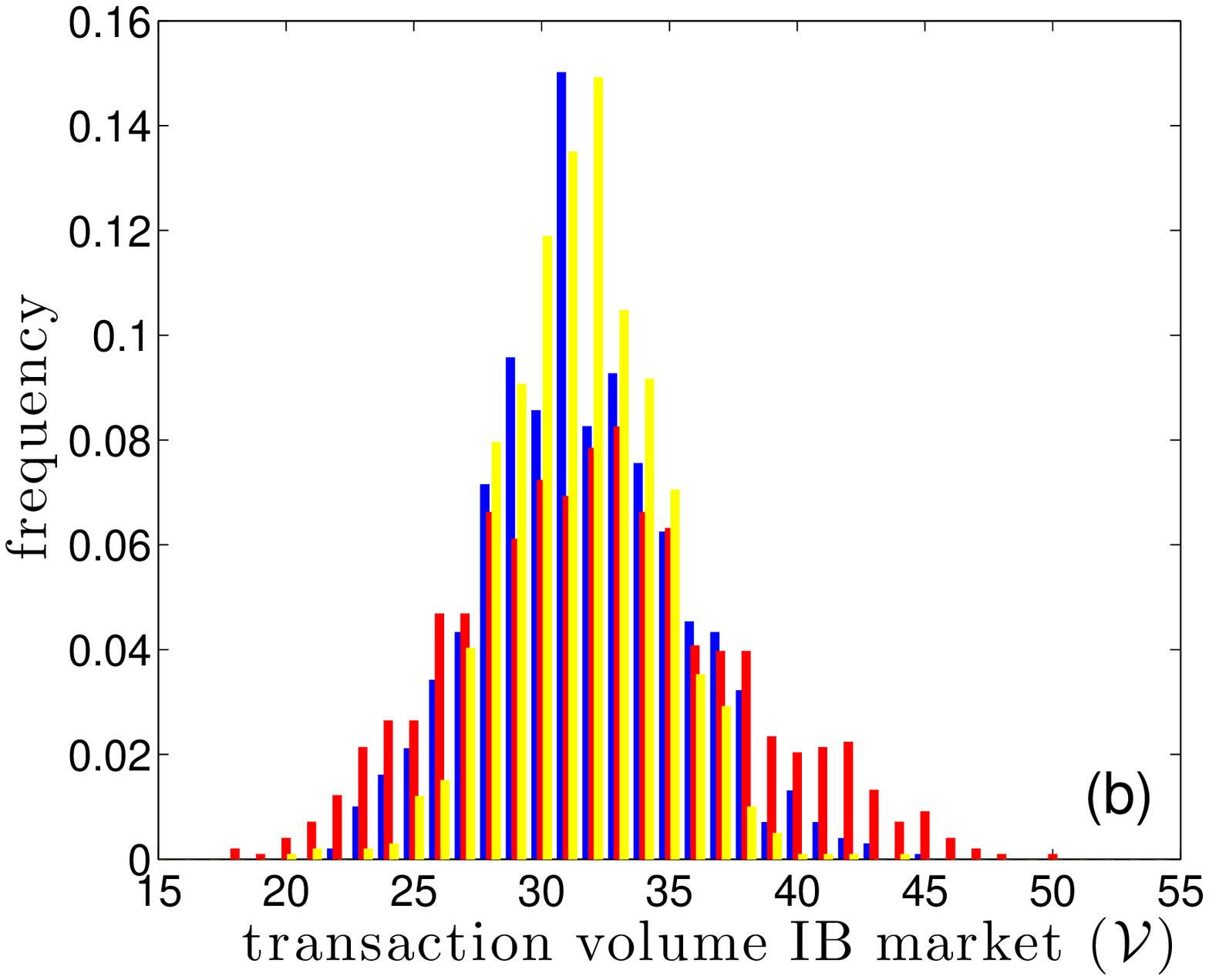}
		\includegraphics[width=.49\textwidth]{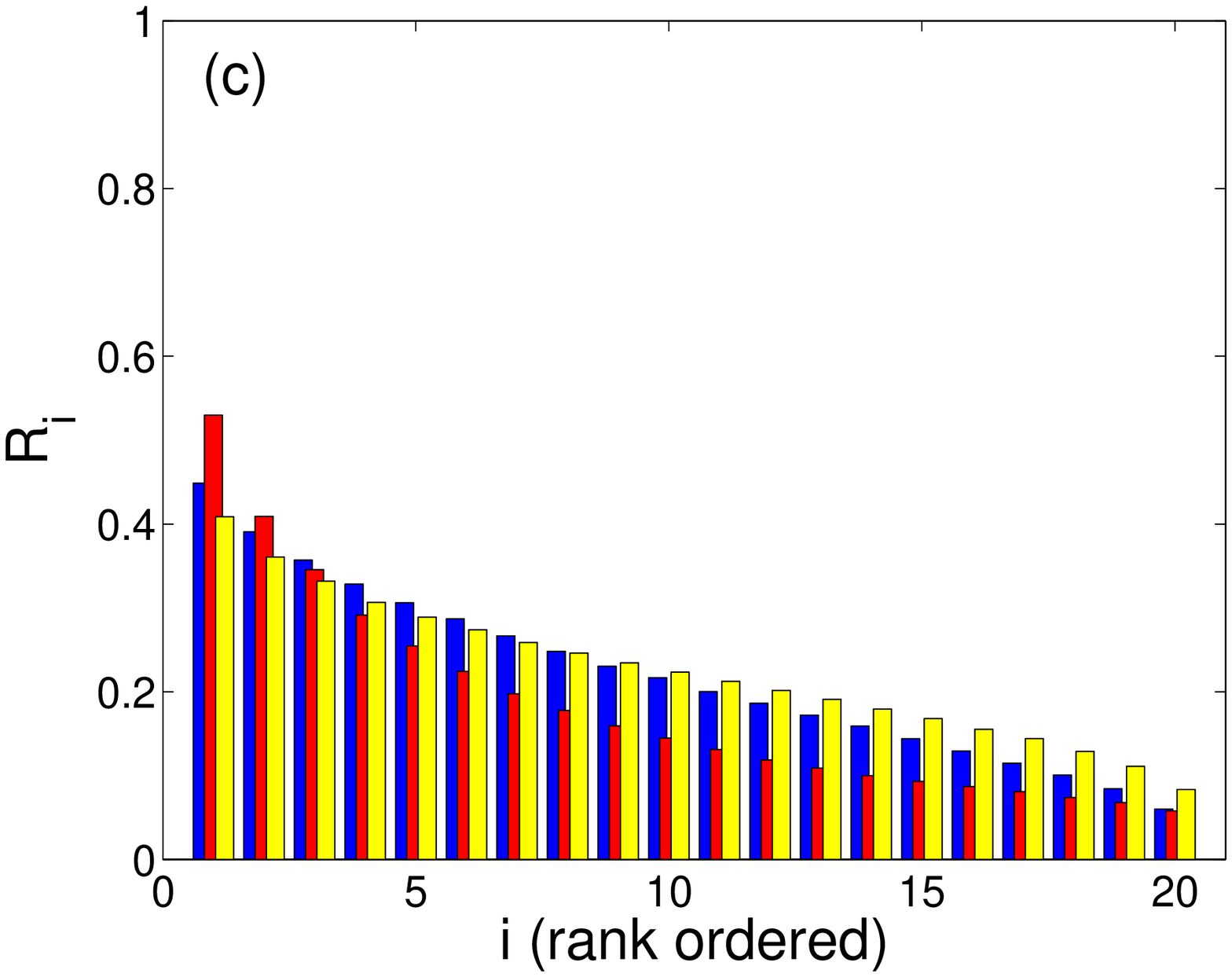}
		\includegraphics[width=.49\textwidth]{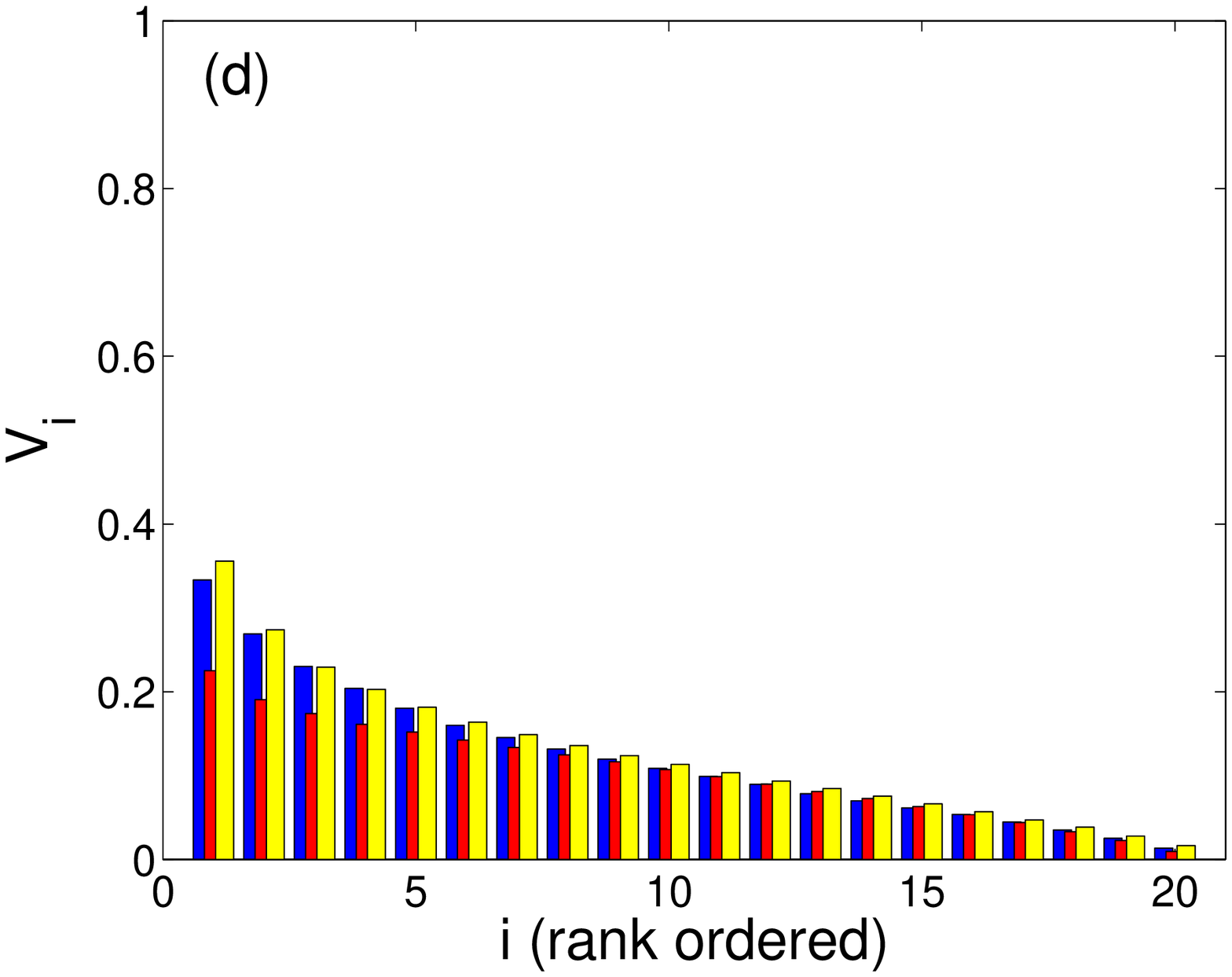}
	\end{center}
	\caption{Comparison of different weight distributions for the individual indicators: weights as specified in Basel III (blue), indicator weight on interbank assets (red) and indicator weight on interbank liabilities (yellow). (a) Distribution of total losses to banks ${\cal L}$, (b) distribution of cascade sizes ${\cal C}$ of defaulting banks, (c) the risk profile for systemic risk given by the distribution of DebtRanks $R_{i}$ (banks are ordered by $R_{i}$, the most systemically important being to the very left), and (d) the risk profile for credit risk shows the distribution of the \emph{average vulnerability} of banks (\cref{vulnerability_section}).} 
	\label{fig:weights}
\end{figure*}
Finally, we study the effects of different weight distributions for the Basel III indicator-based measurement approach. In particular, we are interested in whether it is more effective to have capital surcharges for ``super-spreaders'' or for ``super-vulnerable'' financial institutions. A vulnerable financial institution in this context is an institution that is particularly exposed to failures in the financial system, i.e. has large credit or counterparty risk (CR). Holding assets is generally associated with the risk of losing the value of an investment. Whereas a liability is an obligation towards a counterparty that can have an impact if not fulfilled. To define an indicator that reflects SR and identifies ``super-spreaders'' we set the weight on intra-financial system liabilities to 100\%. For an indicator that reflects CR of interbank assets and identifies vulnerable financial institutions of we set the weight on intra-financial system assets to 100\%. Weights on all other individual indicators are set to zero. Specifically, we consider three different weight distributions. 
\begin{inparaenum}
	[(i)]
	\item weights as specified in Basel III (\cref{indicator}),
	\item indicator weight on interbank liabilities set to 100\%,
	\item indicator weight on interbank assets set to 100\%.
\end{inparaenum}

To illustrate the effects of different weight distributions, we once again use larger capital surcharges for G-SIBs, as specified in Basel III. To allow a comparison between the different weight distributions, we set the level of capital surcharges for each weight distribution to result in a similar level of efficiency (credit volume). Specifically, for the weight distribution as specified in Basel III (\cref{indicator}) we multiply the capital surcharges for G-SIBs as specified in Basel III (\cref{buckets}) by a factor $2.75$, for indicator weight on interbank liabilities by $3$ and for indicator weight on interbank assets by $4.5$. In \cref{fig:weights}(b) we show that efficiency, as measured by transaction volume in the interbank market, is indeed similar for all weight distributions. 

In \cref{fig:weights}(a) we show the losses to banks ${\cal L}$ for the different weight distributions. Clearly, imposing capital surcharges on ``super-spreaders'' (yellow) does indeed reduce the tail of the distribution, but does not, however, get completely rid of large losses in the system (yellow). In contrast, imposing capital surcharges on ``super-vulnerable'' financial institutions (red) shifts the mode of the loss distribution without reducing the tail, thus making medium losses more likely without getting rid of large losses.

To illustrate the effect of the different weight distributions, we show the risk profiles for SR and CR. Risk profiles for SR show the systemic impact as given by the DebtRank $R_{i}$ (banks are ordered by $R_{i}$, the most systemically important being to the very left). Risk profiles for CR show the distribution of \emph{average vulnerability}, i.e. a measure for CR in a financial network, for details see \cref{vulnerability_section}. The corresponding SR- and CR-profiles are shown for the different weight distributions in \cref{fig:weights}(c) and \cref{fig:weights}(d). Imposing capital surcharges on ``super-spreaders'' leads to a more homogeneous spreading of SR across all agents, as shown in \cref{fig:weights}(c) (yellow). Whereas capital surcharges for ``super-vulnerable'' financial institutions leads to a more homogeneous spreading of CR (\cref{fig:weights}(d) (red)). Interestingly, homogeneous spreading of CR (SR) leads to a more unevenly distributed SR (CR), as seen by comparing \cref{fig:weights}(c) and \cref{fig:weights}(d). 

\section{Discussion} \label{discussion}
We use an ABM to study and compare the consequences of the two different options for the regulation of SR. In particular we compare financial regulation that attempts to reduce the financial fragility of SIFIs (Basel III capital surcharges for G-SIBs) with a regulation policy that aims directly at reshaping the topology of financial networks (SRT).

SR emerges in the ABM through two mechanisms, either through interconnectedness of the financial system or through synchronization of the behavior of agents. Cascading failure of banks in the ABM can be explained as follows. Triggered by a default of a firm on a loan, a bank may suffer losses exceeding its absorbance capacity. The bank fails and due to its exposure on the interbank market, other banks may fail as well. Cascading failure can be seen in \cref{fig:regulation_type}(b). 

Basel III capital surcharges for G-SIBs work in the ABM as follows. Demand for commercial loans from firms depends mainly on the expected demand for goods. Firms approach different banks for loan requests. If banks have different capital requirements, banks with lower capital requirements may have a higher leverage and provide more loans. Effectively, different capital requirement produce inhomogeneous leverage levels in the banking system. Imposing capital surcharges for G-SIBs means that banks with a potentially large impact on others must have sizable capital buffers. With capital surcharges for G-SIBs, non-important banks can use higher leverage until they become systemically important themselves. This leads to a more homogenous spreading of SR, albeit that reduction and mitigation of SR is achieved ``indirectly''. 

The SRT leads to a self-organized reduction of SR in the following way \cite{Poledna:2014aa}: Banks looking for credit will try to avoid this tax by looking for credit opportunities that do not increase SR and are thus tax-free. As a result, the financial network rearranges itself toward a topology that, in combination with the financial conditions of individual institutions, will lead to a new topology where cascading failures can no longer occur. The reduction and mitigation of SR is achieved ``directly'' by re-shaping the network.

From the experiments we conclude that Basel III capital surcharges for G-SIBs reduce and mitigate SR at the cost of a loss of efficiency of the financial system (\cref{fig:surcharges}(a) and \cref{fig:surcharges}(c)). This is because overall leverage is reduced by capital surcharges. On the other hand, the SRT keeps the efficiency comparable to the unregulated financial system. However it avoids cascades due to the emergent network structure. This means -- potentially much higher -- capital requirements for ``super-spreaders'' do address SR but at the cost of efficiency. 

SR through synchronization emerges in the ABM in the following way: Similarly to the root cause of cascading failure in the banking system, an initial shock from losses from commercial loans can lead to a synchronization of firm defaults. If banks are able to absorb initial losses, the losses still result in an implicit increase in leverage. In order to meet capital requirements (Basel III), the bank needs to reduce the volume of commercial loans. This creates a feedback effect of firms suffering from reduced liquidity, which in turn increases the probability of defaults and closes the cycle. This feedback effect is inherent to the system and can be re-inforced by Basel III capital surcharges for G-SIBs. In the experiments we see that this feedback effect gets more pronounced with larger capital surcharges for G-SIBs (\cref{fig:surcharges}(b)). Here we see an increase in cascading failure of firms with increasing capital surcharges for G-SIBs. This means capital surcharges for G-SIBs can have pro-cyclical side effects.

Basel III measures systemic importance with an indicator-based measurement approach. The indicator approach is based on several individual indicators consisting of banks' assets and liabilities. By studying the effect of different weight distributions for the individual indicators, we find that Basel III capital surcharges for G-SIBs are more effective with higher weights on liabilities (\cref{fig:weights}(a)). This means -- quite intuitively -- that asset-based indicators are more suitable for controlling credit risk and liability based indicators are more suitable for controlling systemic risk. Since the indicators in the Basel proposal are predominantly based on assets, the Basel III indicator-based measurement approach captures predominantly credit risk and therefore does not meet it's declared objective.

To conclude with a summary at a very high level, the policy implications obtained with the ABM are  
\begin{inparaenum}
	[(i)]
	\item re-shaping financial networks is more effective and efficient than reducing leverage,
	\item capital surcharges for G-SIBs (``super-spreaders'') can reduce SR,
	\item but must be larger than specified in Basel III to have an measurable impact,
	\item and thus cause a loss of efficiency.
	\item Basel III capital surcharges for G-SIBs can have pro-cyclical side effects.
\end{inparaenum} 

\section{Methods} \label{methods}
\subsection{DebtRank} \label{debtrank_section} DebtRank is a recursive method suggested in \citet{Battiston:2012aa} to determine the systemic relevance of nodes in financial networks. It is a number measuring the fraction of the total economic value in the network that is potentially affected by a node or a set of nodes. $L_{ij}$ denotes the interbank liability network at any given moment (loans of bank $j$ to bank $i$), and $C_{i}$ is the capital of bank $i$. If bank $i$ defaults and cannot repay its loans, bank $j$ loses the loans $L_{ij}$. If $j$ does not have enough capital available to cover the loss, $j$ also defaults. The impact of bank $i$ on bank $j$ (in case of a default of $i$) is therefore defined as 
\begin{equation}
	\label{impact} W_{ij} = \min \left[1,\frac{L_{ij}}{C_{j}} \right] \quad. 
\end{equation}
The value of the impact of bank $i$ on its neighbors is $I_{i} = \sum_{j} W_{ij} v_{j}$. The impact is measured by the {\em economic value} $v_{i}$ of bank $i$. For the economic value we use two different proxies. Given the total outstanding interbank exposures of bank $i$, $L_{i}=\sum_{j}L_{ji}$, its economic value is defined as 
\begin{equation}
	\label{ecovalue} v_{i}=L_{i}/\sum_{j}L_{j} \quad. 
\end{equation}
To take into account the impact of nodes at distance two and higher, this has to be computed recursively. If the network $W_{ij}$ contains cycles, the impact can exceed one. To avoid this problem an alternative was suggested in \citet{Battiston:2012aa}, where two state variables, $h_{\rm i}(t)$ and $s_{\rm i}(t)$, are assigned to each node. $h_{\rm i}$ is a continuous variable between zero and one; $s_{\rm i}$ is a discrete state variable for three possible states, undistressed, distressed, and inactive, $s_{\rm i} \in \{U, D, I\}$. The initial conditions are $h_{i}(1) = \Psi \, , \forall i \in S ;\; h_{i}(1)=0 \, , \forall i \not \in S$, and $s_{i}(1) = D \, , \forall i \in S ;\; s_{i}(1) = U \, , \forall i \not \in S$ (parameter $\Psi$ quantifies the initial level of distress: $\Psi \in [0, 1]$, with $\Psi = 1$ meaning default). The dynamics of $h_i$ is then specified by 
\begin{equation}
	h_{i}(t) = \min\left[1,h_{i}(t-1)+\sum_{j\mid s_{j}(t-1) = D} W_{ ji}h_{j}(t-1) \right] \quad. 
\end{equation}
The sum extends over these $j$, for which $s_{j}(t-1) = D$, 
\begin{equation}
	s_{i}(t) = 
	\begin{cases}
		D & \text{if } h_{i}(t) > 0; s_{i}(t-1) \neq I ,\\
		I & \text{if } s_{i}(t-1) = D , \\
		s_{i}(t-1) & \text{otherwise} \quad. 
	\end{cases}
\end{equation}
The DebtRank of the set $S$ (set of nodes in distress at time $1$), is $R^{\prime}_S = \sum_{j} h_{j}(T)v_{j} - \sum_{j} h_{j}(1)v_{j}$, and measures the distress in the system, excluding the initial distress. If $S$ is a single node, the DebtRank measures its systemic impact on the network. The DebtRank of $S$ containing only the single node $i$ is 
\begin{equation}
	\label{debtrank} R^{\prime}_{i} = \sum_{j} h_{j}(T)v_{j} - h_{i}(1)v_{i} \quad. 
\end{equation}
The DebtRank, as defined in \cref{debtrank}, excludes the loss generated directly by the default of the node itself and measures only the impact on the rest of the system through default contagion. For some purposes, however, it is useful to include the direct loss of a default of $i$ as well. The total loss caused by the set of nodes $S$ in distress at time $1$, including the initial distress is 
\begin{equation}
	\label{debtrank_self} R_S = \sum_{j} h_{j}(T)v_{j} \quad. 
\end{equation}

\subsection{Expected systemic loss}
The precise meaning of the DebtRank allows us to define the \emph{expected systemic loss} for the entire economy \citep{Poledna:2015aa,Poledna:2014aa}. Assuming that we have $B$ banks in the system, the expected systemic loss can be approximated by 
\begin{equation}
	EL^{\rm syst}(t) = V(t) \sum_{i=1}^{B} p_i(t) R_i(t) \quad, \label{EL}
\end{equation}
with $p_i(t)$ the probability of default of node $i$, and $V(t)$ the combined economic value of all nodes at time $t$. For details and the derivation, see \citep{Poledna:2015aa}. 

To calculate the marginal contributions to the expected systemic loss, we start by defining the \emph{net liability network} $L^{\rm net}_{ij}(t)=\max[0,L_{ij}(t)-L_{ji}(t)]$. After we add a specific liability $L_{mn}(t)$, we denote the liability network by 
\begin{equation}
	L^{(+mn)}_{ij}(t) = L^{\rm net}_{ij}(t) + \sum_{m,n}\delta_{im}\delta_{jn} L_{mn}(t) \quad, 
\end{equation}
where $\delta_{ij}$ is the Kronecker symbol. The marginal contribution of the specific liability $L_{mn}(t)$ on the expected systemic loss is 
\begin{multline}
	\Delta^{(+mn)} EL^{\rm syst}(t) = \\
	= \sum_{i=1}^{B} p_i(t) \left(V^{(+mn)}(t)R^{(+mn)}_i(t) - V(t)R_i(t) \right) \quad , \label{marginal_effect}
\end{multline}
where $R^{(+mn)}_i(t)=R_i(L^{(+mn)}_{ij}(t),C_i(t))$ is the DebtRank of the liability network and $V^{(+mn)}(t)$ the total economic value with the added liability $L_{mn}(t)$. Clearly, a positive $\Delta^{(+mn)} EL^{\rm syst}(t)$ means that $L_{mn}(t)$ increases the total SR. 

Finally, the marginal contribution of a single loan (or a transaction leading to that loan) can be calculated. We denote a loan of bank $i$ to bank $j$ by $l_{ijk}$. The liability network changes to 
\begin{equation}
	L^{(+k)}_{ij}(t) = L^{\rm net}_{ij}(t) + \sum_{m,n,k} \delta_{im}\delta_{jn} \delta_{kk} l_{mnk}(t) \quad. \label{Lwithoutloan} 
\end{equation}
Since $i$ and $j$ can have a number of loans at a given time $t$, the index $k$ numbers a specific loan between $i$ and $j$. The marginal contribution of a single loan (transaction) $\Delta^{(+k)} EL^{\rm syst}(t)$, is obtained by substituting $L^{(+mn)}_{ij}(t)$ by $L^{(+k)}_{ij}(t)$ in \cref{marginal_effect}. In this way every existing loan in the financial system, as well as every hypothetical one, can be evaluated with respect to its marginal contribution to overall SR.

\subsection{Systemic risk tax} \label{srt}
The central idea of the SRT is to tax every transaction between any two counterparties that increases SR in the system \citep{Poledna:2014aa}. The size of the tax is proportional to the increase of the expected systemic loss that this transaction adds to the system as seen at time $t$. The SRT for a transaction $l_{ijk}(t)$ between two banks $i$ and $j$ is given by
\begin{multline}
	SRT_{ij}^{(+k)}(t) = \\
	= \zeta \max \left[0, \sum_i p_i(t) \left(V^{(+k)}(t)R^{(+k)}_i(t) - V(t)R_i(t) \right) \right] \quad. \label{srteq_simple} 
\end{multline}
$\zeta$ is a proportionality constant that specifies how much of the generated expected systemic loss is taxed. $\zeta=1$ means that 100\% of the expected systemic loss will be charged. $\zeta<1$ means that only a fraction of the true SR increase is added on to the tax due from the institution responsible. For details, see \citep{Poledna:2014aa}. 

\subsection{Average vulnerability} \label{vulnerability_section} Using DebtRank we can also define an \emph{average vulnerability} of a node in the liability network $L_{ij}$ \cite{Aoyama:2013aa}. Recall that $h_{i}(T)$ is the state variable, which describes the health of a node in terms of equity capital after $T$ time steps ($s_{\rm i} \in \{U, I\}$, i.e. all nodes are either undistressed or inactive). When calculating $h_{i}(T)$ for $S_f$ containing only a single node $i$, we can define $h_{ij}(T)$ as the health of bank $j$ in case of default of bank $i$ 
\begin{equation}
	\label{hmatrix_eq} h_{ij}(T) = 
	\begin{bmatrix}
		h_{1}(T) \dots h_{N}(T) 
	\end{bmatrix}
	\quad. 
\end{equation}
Note that is necessary to simulate the default of every single node of the liability network $L_{ij}$ separately to obtain $h_{ij}$. With $h_{ij}$ we can define the average vulnerability of a node in the liability network as the average impact of other nodes on node $i$
\begin{equation}
	V_i = \frac{1}{B}\sum_j h_{ji} \quad,
\end{equation} where $B$ is the number of nodes in the liability network $L_{ij}$. 

\subsection{Measures for losses, default cascades and transaction volume} \label{risk_measures} We use the following three observables: (1) the size of the cascade, ${\cal C}$ as the number of defaulting banks triggered by an initial bank default ($1\leq {\cal C} \leq B$), (2) the total losses to banks following a default or cascade of defaults, ${\cal L}= \sum_{i \in I}\sum_{j=1}^B L_{ ij}(t)$, where $I$ is the set of defaulting banks, and (3) the average transaction volume in the interbank market in simulation runs longer than $100$ time steps, 
\begin{equation}
	\mathcal{V}=\frac{1}{T}\sum_{t=1}^T\sum_{j=1}^B\sum_{i=1}^B \sum_{k \in K}l_{jik}(t) \quad, 
\end{equation}
where $K$ represents new interbank loans at time step $t$.

\end{document}